\documentclass[12pt]{iopart}
\usepackage{graphicx}

\usepackage{cite}
\usepackage{iopams} 
\usepackage{amsfonts}
\usepackage{xcolor}

\expandafter\let\csname equation*\endcsname\relax
\expandafter\let\csname endequation*\endcsname\relax
\usepackage{amsmath}
\usepackage{mathtools}
\usepackage{caption}
\usepackage{subcaption}

\begin{document}

\title{Effect of collisions on the plasma sheath in the presence of an inhomogeneous magnetic field}

\author{K. Deka$^{1}$, S. Adhikari$^{2}$, R. Moulick$^{3}$, S. S. Kausik$^{1,a)}$, and B. K. Saikia$^{1}$}

\address{$^1$Centre of Plasma Physics-Institute for Plasma Research, Sonapur 782402, Assam, India}

\address{$^2$Department of Physics, University of Oslo, PO Box 1048 Blindern, NO-0316 Oslo,
Norway}
\address{$^3$Department of Physics, Rangapara College, Rangapara, Sonitpur, Assam - 784505 (India)}

\vspace{10pt}
\address{$^{a)}$Email: kausikss@rediffmail.com}

\begin{abstract}
A low-pressure magnetized plasma is studied to find the dependency of sheath properties on ion-neutral collisions in presence of an inhomogeneous magnetic field. A self-consistent one-dimensional two-fluid hydrodynamic model is considered, and the system of equations is solved numerically. The study reveals that the width of the plasma sheath expands and space charge increases with collisions. The ion-neutral collisions and the inhomogeneous magnetic field restrict the ions to move towards the surface. The movement of the ions towards the wall can be controlled by choosing a suitable configuration of the magnetic field and ion-neutral collision frequency. A comparison between two different magnetic field configuration has been presented along side to differentiate the commonly found scenarios in the field. The outcome of the study is supposed to help in understanding the complex dynamics of ions in plasma confinement and plasma processing of materials. Furthermore, the present work seeks to create a framework for two-fluid modeling of magnetized plasmas with any arbitrary magnetic field profiles. The analysis provided here is supposed to act as a basis for any future work in the respective field.

\end{abstract}

\maketitle
\section{\label{sec:level1}Introduction}

The plasma sheath is a very old area of research in plasma physics. It remains to be one of the important fields due to its crucial role in many plasma applications such as plasma processing, fabrication of semiconductor devices, etching, etc. Moreover, the study of this space charge layer has gathered a lot of attention due to the growing importance of fusion reactors {\cite{masoudi2015, Adhikari2018}}. In magnetically confined devices like tokamak, particle transport in the edge region often takes place in the presence of an inhomogeneous magnetic field. Therefore, such applications strive many researchers to study the behavior of the sheath region and its influence on particle transport.
The properties of the plasma sheath have been studied for both electrostatic and magnetized environment {\cite{moulick2019}}. However, for the magnetized case, most of the studies have been carried out in the presence of a uniform oblique magnetic field {\cite{pandey2008,adhikari2017,chodura1982, riemann1994}}. A brief literature survey of the field is incorporated in the following section to portray the present status.


In presence of an external magnetic field, it is seen that the dynamics of the sheath formation is altered{\cite{masoudi2015,cohen1995, chodura1982}}. The ion-neutral collision also plays an important role in the formation and stability of sheath structure. The collisional plasma sheath has been investigated by many researchers{\cite{basnet2018, moulick2019, ou2012, masoudi2007,sharma2020effect}}. It has been observed that the collisional force affects the distribution of ion density and drift velocity {\cite{{basnet2018,moulick2013,valentini2000}}}. The study of collisional plasma sheath has shown that collision alters the condition for the formation of the sheath. In the presence of collision, the sheath may be formed with ions having velocity much less than the ion-acoustic speed, violating the Bohm's criterion for an  electrostatic sheath {\cite{moulick2019,moulick2014,bohm1991}}. Such a scenario has been explored by many authors, and many attempts have been made to find the right criterion for the sheath formation in the presence of collision. Considering a two-fluid model Valentini{\cite{valentini1996}}, studied the effect of elastic collisions and charge exchange on sheath criterion analytically. Moulick \textit{et al.} {\cite{moulick2017criterion}} derived the sheath criterion that included the effects of a magnetic field as well as collision. It reduces to electrostatic sheath criterion when collision and magnetic field are absent. Riemann studied the collision dominated boundary layer using kinetic approach {\cite{riemann1981}} in which the charge exchange collisions with the neutrals governs the ion dynamics. The study revealed that the ion distribution function resembles that of a half Maxwellian inside the plasma sheath.  The thickness of magnetized presheath is also found to be angle-dependent, in a low-pressure collisional environment {\cite{moulick2019}}. The magnetic pre-sheath ceases to exist with the increase in the angle of incidence. Hatami \textit{et al.} {\cite{hatami2008}} carried out a similar study on the collisional effects in magnetized plasma sheath with two species of positive ions. The findings of this study became particularly important as the kinetic energies of both ion species were found to decrease with the increase in ion-neutral collision frequency.


Collisional plasma sheath is usually modeled in two different ways{\cite{moulick2019,franklin2001}}.
\begin{itemize}
    \item Constant collision frequency scheme. It assumes that the collision frequency does not depend upon the ion (or electron) velocity.
     \item The constant collision cross-section scheme, where the collision frequency varies with the ion (or electron) velocity (fluid velocity).

\end{itemize}
In the present study, the constant collision frequency scheme is used to model the electron-neutral collisions{\cite{masoudi2009}} $({\nu}_{en})$. On the other hand, constant collision cross-section model is adopted for the ion-neutral collisions{\cite{moulick2019, hatami2008}} $({\nu}_{in})$.

 An extensive amount of study has been accomplished on plasma sheath in the presence of a uniform and oblique magnetic field (UOM) as the field structure is quite relevant to the fusion environment {\cite{moulick2019, gyergyek2016, adhikari2017, masoudi2015, Adhikari2018}}. In particular, such studies under the influence of collision have become a trend in recent times. In a recent study {\cite{thakur2018}},  fluid analysis of plasma sheath in a cylindrical geometry has shown that the collisional force dominates over the effect of the magnetic field on ion velocity. It deals with the properties of cylindrical plasma, which is common in the laboratory. Pandey {\textit{et al.}} {\cite{pandey2008}} investigated the structure of the plasma sheath in the presence of an oblique magnetic field using a two-fluid model. It focuses on the impact of plasma magnetization, plasma ionization and collisions on the plasma sheath for different orientations of the magnetic field. Holland {\textit{et al.}} {\cite{holland1993}} explored the boundary layer of plasma in the presence of tilted magnetic field $\theta<9^{\circ}$ using a one-dimensional time-dependent model, in which they consider the electrons as fluid and ions as Maxwellian. The results reveal that the properties of the plasma sheath are determined by the ratio of convective and diffusive electron flows. Another important aspect of the grazing angle incidence is, it helps in reducing the impact of energetic ion-bombardment on the surface. Devaux and Manfredi {\cite{devaux2008}} established the theory using a kinetic model. Adhikari \textit{et al.} {\cite{adhikari2017}}, studied the influence of force fields on ion-dynamics inside a magnetized plasma sheath. The role of Lorentz force and energy acquired by the ions on their way towards the surface is explored in the paper. Such an analysis of the force helps understand the kinetics of the ions and hence various sheath properties. The plasma sheath is also studied under the influence of uniform magnetic field parallel to the surface {\cite{natalia2010}}. Such a field configuration is helpful to confine a plasma as it restricts the mobility of both the electrons and ions to some extent. In stellarator or tokamak devices, important sections that come in contact with the plasma, are nearly parallel to the magnetic field. Hence, the understanding of the plasma sheath in such an environment is quite helpful in determining the boundary conditions to aid the simulation of the scrape of layer (SOL) {\cite{moritz2016}}. The particle simulation of a plasma sheath in the presence of a  parallel magnetic field was studied by Daube {\textit{et al.}} {\cite{daube1998}}, in which the particle transport towards the surface is provided by charge exchange collisions with a neutral background. They got a poor agreement in the velocity distribution of ions obtained from the particle in cell method and Boltzmann equation which was attributed to the time-dependent fluctuations of the electric field. A two-fluid model of Plasma Wall Transition (PWT) layer in presence of a parallel field was investigated by Tskhakaya \textit{et al.} {\cite{tskhakaya2005}}. It has shown that firstly the ionization and recombination are two vital processes to join the PWT layer with bulk plasma, and secondly the Boltzmann distribution of electrons in magnetized plasma is suspicious.  The kinetic simulation of a plasma sheath for parallel field configuration was carried out by Li and Wang {\cite{li2018}}. The simulation throws light on the following two facts, firstly the spatial structure of the sheath depends on ion Debye length contrary to ion Larmor radius and electron Debye length, and secondly time for the formation of the sheath is of the order of ion-cyclotron time. 
     
Although this significant volume of literature has provided a detailed account of the plasma sheath in the presence of UOM and magnetic field parallel to the surface, the field lacks findings on spatially varying magnetic field. In this article, we have concentrated on the sheath formation near a wall in the presence of an inhomogeneous magnetic field. The motivation lying behind the present work is to understand the nature of the plasma sheath in a collisional environment in the presence of an inhomogeneous magnetic field, which might be of use in material processing as well as in magnetically confined plasma devices. In the rest of the paper the inhomogeneous magnetic field is represented in short by IHM for convenience.

 The present work is arranged in six sections. In section II, the basic equations and the theoretical model is discussed. In section III, the normalization of the variables is done.  Section IV takes into account the required numerical scheme for the study. Section V includes the results and discussion. The conclusion of the work is presented in section VI.  An 'Appendix' section describing the magnetic field profile related to the present work can be found at the end.

\section{Theoretical Model and basic equations}

 As mentioned earlier most of the study of plasma sheath has been carried out in the presence of an uniform magnetic field. The properties of the plasma sheath is well established in the presence of the such magnetic field configuration. But there is no such study on the plasma sheath dealing with a magnetic field varying spatially along the system length \textit{i.e} in the presence of an inhomogeneous magnetic field. The simplest form of an inhomogeneous magnetic field in free space is given by{\cite{bittencourt}},

\begin{equation}
{B=B_0(1+{\alpha}z)\hat{i}+(B_0{{\alpha}x})\hat{k}}
\end{equation}

 Where ${\alpha}$ is a constant having dimension of the inverse of length.  The first term of equation (1) is called the gradient term. The second term is related to the curvature of the field lines and is called the curvature term. In the above equation, $\hat{i}$ and $\hat{k}$ represent the unit vectors along x and z axes, respectively. The presence of the curvature term along with the gradient term makes the problem two-dimensional thereby increases the complexity of the problem.
 
 The strength of the magnetic field is given by
 
\begin{equation*}
B=B_0\sqrt{(1+\alpha{z})^2+({\alpha}x)^2}
\end{equation*}
 
  For the present study, considering both the terms of the magnetic field, the strength of the total magnetic field varies from 0.5T to 0.79T. On the other hand the strength of the gradient term alone for the present work varies from 0.5T to 0.75T. It can be seen that total field strength is comparable with the field strength of the gradient term of the magnetic field in particular. Initially the magnetic field represented by equation (1) is parallel to the wall and gradually the field makes a very small inclination with the wall. This inclination slowly increases towards the wall (At the wall $\theta<19^{0}$). Further, if we confine our study in a region near the z-axis of the x-z plane, the magnitude of ${\theta}$ will even be smaller and the field aligns itself almost parallel to the wall. Therefore, it can be assumed that the effect of the gradient term will dominate over the curvature term of the magnetic field. This assumption enables the 2D problem to be treated as an 1D problem. Such approximation is not entirely unfamiliar. In a similar way, Ebersohn \textit{et al.}{\cite{ebersohn2017}} reduced the 2D problem of plasma flow in a strong external magnetic field to 1D problem by considering only the axial variation of the magnetic field and assuming that the plasma properties are constant across the flux tube cross-section.

 In fluid theory, the magnetic field acting on the charged particles appears through the momentum equation for each species. It can be seen that for a 1D sheath, formed along the z-axis, the force due to the curvature term does not appear in the momentum equations for both the ions and electrons. So, the curvature term does not affect the sheath formation along z-axis despite its presence in the original field profile. The necessary calculations are shown in the Appendix. Hence the effective magnetic field for the present study can be taken as,

  \begin{equation}
{B=B_0(1+{\alpha}z)\hat{i}}
\end{equation}
The magnetic field is along the x-axis, and it increases linearly towards the surface.

\begin{figure}
\centering
\includegraphics[width=12cm]{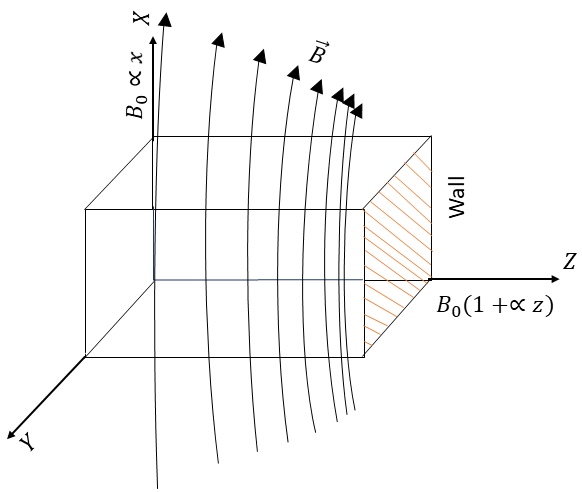}
\caption{ Schematic diagram of the theoretical model}
\label{Fig1}
\end{figure}

 In the present study a steady-state low-pressure $({T_i<<T_e})$ plasma consisting of positive ions and electrons is considered. A schematic diagram of the theoretical model is shown in Fig. {\ref{Fig1}}. The problem has a perspective, where, a magnetic field is aligned almost parallel to the surface in order to restrict the plasma coming directly in contact with the surface. In the theoretical model, it is assumed that the z-axis is perpendicular to the surface and the sheath parameters, such as plasma density, potential, electric field, etc. are varying along z-axis {\cite{hatami2012,adhikari2017,riemann1994theory}}. An expression for the field has been provided in Equation (2). The reason behind such a choice of magnetic field is pretty straight forward. The physics of sheath is already a complex topic. Our aim is to introduce the effect of this IHM in the intricate sheath problem with minimal complexity. 
 
The positive ions and the electrons are described by the fluid equations{\cite{ghomi2010, Franklin1970}}. The momentum equation for the ions and electrons are expressed as,
\begin{equation}
m_in_iv_{iz}\frac{d\vec{v_i}}{dz}=-n_ie\frac{\partial\phi}{\partial{z}}\hat{k}+n_ie(\vec{v_i}\times \vec{B})-m_i\vec{v_i}S_i-m_in_i\vec{v_i}\nu_{in}
\end{equation}
\begin{equation}
m_en_ev_{ez}\frac{d\vec{v_e}}{dz}=n_ee\frac{\partial\phi}{\partial{z}}\hat{k}-n_ee(\vec{v_e}\times\vec{B})-m_e\vec{v_e}S_i-{T_e}\frac{d{n_e}}{dz}\hat{k}-m_en_e\vec{v_e}\nu_{en} \end{equation}
Where $S_i$ represents source term which is given by,

\begin{equation}
S_i={n}_{e}\mathbb{Z}  \end{equation}
Here, $\hat{k}$, $\phi$, $n_i$, $n_e$, $\nu_{in}$ and $\nu_{en}$ represent unit vector along the z-axis, electric potential, ion-density, electron density, ion-neutral and electron-neutral collision frequency respectively. $\mathbb{Z}$ in the source term stands for the ionization frequency of the plasma. The rest of the symbols have their usual meaning. The ions are considered to be cold and singly charged.

The velocity components of the ions are $v_{ix}$, $v_{iy}$, and $v_{iz}$ along $x$, $y$, and $z$ axes, respectively.  Accordingly, the momentum equation can be resolved into three components.

\begin{equation}
v_{iz}\frac{dv_{ix}}{dz}=-\bigg(\frac{n_e}{n_i}\bigg)v_{ix}{\mathbb{Z}}-v_{ix}\nu_{in}
\end{equation}
\begin{equation}
 v_{iz}\frac{dv_{iy}}{dz}=\bigg(\frac{eB}{m_i}\bigg)v_{iz}-\bigg(\frac{n_e}{n_i}\bigg)v_{iy}{\mathbb{Z}}-v_{iy}\nu_{in}
\end{equation}
\begin{equation}
v_{iz}\frac{dv_{iz}}{dz}=-\frac{e}{m_i}\bigg(\frac{\partial\phi}{\partial{z}}\bigg)-\bigg(\frac{eB}{m_i}\bigg)v_{iy}-\bigg(\frac{n_e}{n_i}\bigg)v_{iz}{\mathbb{Z}}-v_{iz}\nu_{in}
\end{equation}

The components of the momentum equation for the electrons are given by
\begin{equation}
 v_{ez}\frac{dv_{ex}}{dz}=-{\mathbb{Z}}v_{ex}-v_{ex}\nu_{en}
\end{equation}
\begin{equation}
v_{ez}\frac{dv_{ey}}{dz}=-\bigg(\frac{eB}{m_e}\bigg)v_{ez}-{\mathbb{Z}}v_{ey}-v_{ey}\nu_{en}
\end{equation}
\begin{equation}
v_{ez}\frac{dv_{ez}}{dz}=\frac{e}{m_e}\bigg(\frac{\partial\phi}{\partial{z}}\bigg)+\bigg(\frac{eB}{m_e}\bigg)v_{ey}-\bigg(\frac{T_e}{{n_e}{m_e}}\bigg)\frac{d{n_e}}{dz}-{\mathbb{Z}}v_{ez}-v_{ez}\nu_{en}
\end{equation}
The source term $S_i$ accounts for the difference between the number of ions that are created and annihilated per unit volume per unit time in the plasma {\cite{gyergyek2016,robertson2013}}. Considering the source term, the continuity equations for the ions and electrons takes the form,
\begin{equation}
  \frac{\partial}{\partial{z}}(n_iv_{iz})=S_i
\end{equation}
\begin{equation}
{\frac{\partial}{\partial{z}}({n_e}{v_{ez}})=S_i}
\end{equation}

There are usually three different forms of source term found in the literature,
\begin{itemize}
\item    Zero source term \textit{i.e.} the electrons and ions are neither produced nor lost throughout the plasma.

   \begin{equation*}{ S_i=0}\end{equation*} 

\item    Constant source term, \begin{equation*}{ S_i=\frac{n_0}{\tau}}\end{equation*}
       Where ${\tau}$ is called ionization time.                     
\item    Exponential source term{\cite{gyergyek2016}}, 
\begin{equation*}
S_i={n}_e{\mathbb{Z}}={n_0}\exp\bigg(\frac{e\phi}{kT_e}\bigg){\mathbb{Z}}
\end{equation*}
\end{itemize}

In this work, 

\begin{equation*}
S_i={n}_e{\mathbb{Z}}
\end{equation*}
has been taken as the source term, as the electrons are not considered as Boltzmann distributed.

Finally, the set of equations is closed by the Poisson's equation,

 \begin{equation}
    \frac{{\partial}^2{\phi}}{\partial{{z}^2}}=-\frac{e}{\epsilon_0}(n_i-n_e) 
  \end{equation}
 Here, ${\epsilon}_0$ is the permittivity of free space.
 
\section{Normalized parameters and the scale of simulation}
To solve the basic set of equations numerically, the physical variables have been normalized with the help of the parameters given below, 
\begin{equation*}
 {u_i=\frac{v_{ix}}{c_s}},~{v_i=\frac{v_{iy}}{c_s}},~{w_i=\frac{v_{iz}}{c_s}}
 \end{equation*}
 \begin{equation*}
{{u_e=\frac{v_{ex}}{c_s}},~{v_e=\frac{v_{ey}}{c_s}},~{w_e=\frac{v_{ez}}{c_s}}}
 \end{equation*}
\begin{equation*}
{\eta=\frac{e\phi}{kT_e}},~{\xi=\frac{z}{L}},
~{L=\lambda_{ni}=\frac{c_s}{\mathbb{Z}}}\end{equation*}
\begin{equation*}
{N_i=\frac{n_i}{n_0}},\hspace{5mm}{c_s=\sqrt{\frac{\textit{k}T_e}{m_i}}},
\hspace{5mm}{N_e=\frac{n_e}{n_0}}
\end{equation*}
Here, $\lambda_{ni}$ represents ionization length and $c_s$ is the ion-acoustic speed. After normalization, the equations (6-14) take the following form,

\begin{equation}
\frac{du_i}{d\xi}=-\bigg(\frac{L\mathbb{Z}}{c_s}\  \bigg)\bigg(\frac{N_e}{N_i}\bigg)\bigg(\frac{u_i}{w_i}\bigg)-\bigg(\frac{L\nu_{in}}{c_s}\bigg)\bigg(\frac{u_i}{w_i}\bigg)\end{equation}
\begin{equation}\frac{dv_i}{d\xi}={\gamma}_i-\bigg(\frac{L\mathbb{Z}}{c_s}\  \bigg)\bigg(\frac{N_e}{N_i}\bigg)\bigg(\frac{v_i}{w_i}\bigg)-\bigg(\frac{L\nu_{in}}{c_s}\bigg)\bigg(\frac{v_i}{w_i}\bigg)\end{equation}
\begin{equation}
\frac{dw_i}{d\xi}=-\bigg(\frac{1}{w_i}\bigg)\bigg[\frac{\partial{\eta}}{\partial{\xi}}+{\gamma}_{i}{v_i}+\bigg(\frac{L\nu_{in}}{c_s}\bigg)w_i\bigg]-\bigg(\frac{L\mathbb{Z}}{c_s}\bigg)\bigg(\frac{N_e}{N_i}\bigg)
\end{equation}

\begin{equation}
\frac{dN_i}{d\xi}=\bigg(\frac{N_i}{w_i^2}\bigg)\bigg[\frac{\partial{\eta}}{\partial{\xi}}+{\gamma}_{i}{v_i}+\bigg(\frac{L\nu_{in}}{c_s}\bigg)w_i+2\bigg(\frac{L\mathbb{Z}}{c_s}\bigg)\bigg(\frac{N_e}{N_i}\bigg)w_i\bigg]
\end{equation}

\begin{equation}
{\frac{du_e}{d\xi}=-\bigg(\frac{L\mathbb{Z}}{c_s}\  \bigg)\bigg(\frac{u_e}{w_e}\bigg)-{\delta}\bigg(\frac{u_e}{w_e}\bigg)}\end{equation}

\begin{equation}
\frac{dv_e}{d\xi}=-{\gamma}_e-\bigg(\frac{L\mathbb{Z}}{c_s}\bigg)\bigg(\frac{v_e}{w_e}\bigg)-{\delta}\bigg(\frac{v_e}{w_e}\bigg)
\end{equation}

\begin{equation}
\frac{dw_e}{d\xi}=\bigg(\frac{w_e}{{\mu}w_e^2-1}\bigg)\bigg[\frac{\partial{\eta}}{\partial{\xi}}+{\gamma}_{e}{v_e}{\mu}-{\delta}w_e{\mu}\bigg]-\bigg(\frac{L\mathbb{Z}}{c_s}\bigg)\bigg(\frac{{\mu}w_e^2+1}{{\mu}w_e^2-1}\bigg)
\end{equation}

\begin{equation}
\frac{dN_e}{d\xi}=\bigg(\frac{N_e{\mu}}{{\mu}w_e^2-1}\bigg)\bigg[-\frac{1}{\mu}\frac{\partial{\eta}}{\partial{\xi}}-{\gamma}_{e}{v_e}+{\delta}w_e+2\bigg(\frac{L\mathbb{Z}}{c_s}\bigg)w_e\bigg]
\end{equation}
\begin{equation*}
\begin{multlined}
{where}, \hspace{2mm}
{\gamma}_i=\bigg(\frac{eB}{m_i\mathbb{Z}}\bigg);~~~{\gamma}_e=\bigg(\frac{eB}{m_e\mathbb{Z}}\bigg);\\~~~\delta=\bigg(\frac{L\nu_{en}}{c_s}\bigg);~~~\mu=(m_e/m_i).
\end{multlined}
\end{equation*}

\begin{equation} {\frac{d^2\eta}{d{\xi}^2}=a\bigg({N_e}-{N_i}\bigg)}\end{equation}
 \begin{equation*}~  where,~{a=\bigg(\frac{L}{\lambda_D}\bigg)^2}\end{equation*}
 
The magnetic field can be expressed in terms of normalized distance ${\xi}$, and it takes the form
\begin{equation}
{B=B_0(1+{\alpha}{\lambda}_{ni}\xi)}
\end{equation}
Here, $B_0=0.5~T $ is the value of the magnetic field at ${\xi}=0$. The corresponding value of the magnetic field at the wall is $B_{wall}=0.75~T$. The field strength varies linearly between these two values. 

In this article, the constant collision cross-section is assumed for modelling the ion-neutral collision. The collision parameter for the present model can be defined as {\cite{moulick2019}}
 \begin{equation}
K=\frac{L}{\lambda_i};
 \hspace{5mm}{\nu_{in}=\frac{\mid{v_i}\mid}{\lambda_i}}
\end{equation}
Here, the parameter $\lambda_i$ is the ion mean free path, which is considered constant on a case-by-case basis. $\mid{v_i}\mid$ is the magnitude of the ion fluid velocity.
The equation (25) leads to,
\begin{equation}
\frac{L\nu_{in}}{c_s}=\frac{K\mid{v_i}\mid}{c_s}=K\sqrt{u^2+v^2+w^2}
\end{equation}

If d is the position of the surface for each numerical execution then the normalized coordinate of the surface is ${\xi_w=\big({d}/{L}}\big)$. This makes the graphical visualization easier and provides a new normalized scale between 0 and 1. The domain ${0 \leq{\xi} \leq{\xi_w} }$ after normalization  {\cite{moulick2019, phukan2018}} becomes $0 \leq({{\xi}/{\xi_w}}) \leq{1}$.

\section{Numerical analysis}

The equations (15) - (23) is a set of ordinary differential equations that has been solved numerically using Runge-Kutta integration. 
The initial values for the set of equations, i.e. at the bulk plasma ($\xi=0$) have been evaluated using Taylor's series expansion. In this method, the plasma parameters, such as density, velocity, potential, etc. are expanded in a Taylor series to obtain the coefficients of the series. The method {\cite{moulick2019,forrest1968,edgley1980}} provides precise and accurate initial values to start the numerical calculation. The plasma parameters are expressed in Taylor series as follows,
\begin{equation}
N_p=N_{p_0}+N_{p_1}{\xi}^2+N_{p_2}{\xi}^4+\cdots
\end{equation}
\begin{equation}
V_p=V_{p_1}{\xi}+V_{p_2}{\xi}^3+\cdots
\end{equation}
\begin{equation}
\eta=\eta_1{\xi}^2+\eta_2{\xi}^4+\cdots
\end{equation}

${N_p}$ and  ${V_p} $   depict the density, and velocity for each species in the plasma, and $\eta$  represents plasma potential. The equations (27) - (29) are substituted back in equations (15) - (23) to obtain the coefficients of each term in the series. Using these coefficients, initial values at the first boundary (\textit{i.e} at a point slightly away from the bulk plasma towards the wall) are calculated. These boundary values are used for solving the system of equations numerically. In the present work the MATLAB routine, {{ode45}} has been used. The numerical simulation is terminated at the point, where the electron current equals the ion current. At this point the electron flux must be equal to the ion flux{\cite{forrest1968}}. 

The following plasma parameters are used for the numerical execution,

~~~~~~~~$T_e=3 eV$, $m_i=40~amu$, $n_0=10^{18} m^{-3}, {\delta}=10^3$

\section{Results and Discussion}

Fig. {2} ({a})  shows the variation of space charge for the collisionless case for magnetic field $B=0$. The profile exactly matches with the space charge profile for the collisionless electrostatic case of Moulick \textit{et .al}{\cite{moulick2019}}. This validates our code and the method of solving the differential equations.

In the present work, an attempt has been made to establish the dependence of plasma sheath properties on ion-neutral collision in the presence of IHM. Hence, it is essential to study the space charge profile to have an exact idea about the sheath width. The onset of space charge may be considered as the sheath entrance point {\cite{rakesh2015}}.  It can be seen from {{Fig. 2}} {(b)}, that the sheath width increases with the increase in ion-neutral collision.  Collisions try to restrict the ions from moving towards the surface, which reduces the shielding of the negative potential at the surface, causing the plasma sheath to expand.  A similar result can be found in the following reference for unmagnetized plasmas {\cite{rakesh2015}}. For a particular value of the collision parameter, the IHM prevents the ion movement towards the surface in a similar fashion. So, it is expected that the field is likely to have an indirect effect on the expansion of the sheath. {{Fig. 2}} {(b)} shows that the space charge increases with the increase in the ion-neutral collision frequency, which is different from the result obtained for the electrostatic case {\cite{valentini2000}}. For the electrostatic case, the space charge density is more for the collisionless environment $(K=1)$. With the increase in collisions, the space charge density reduces, and it again increases in the higher collisional regime.

\begin{figure}[ht]
\begin{subfigure}[b]{0.48\textwidth}
         \centering
         \includegraphics[width=\textwidth]{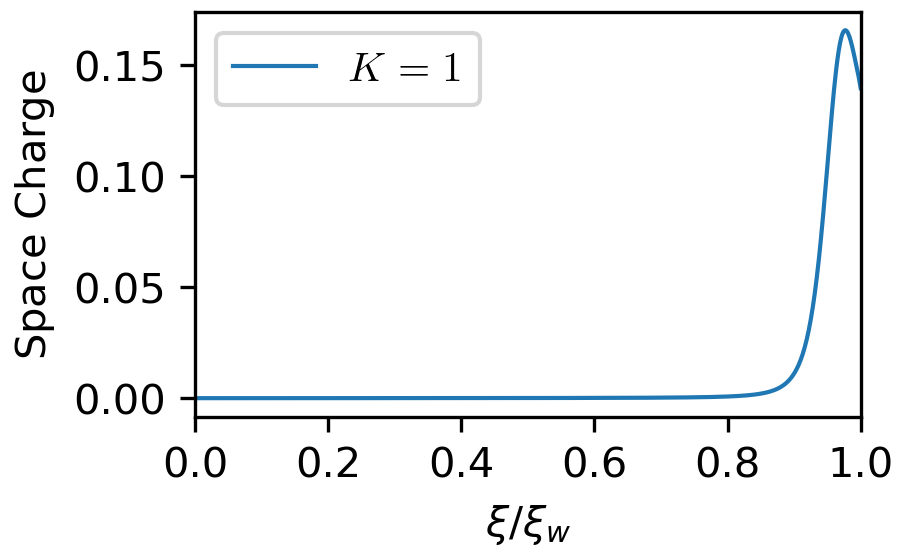}
         \caption{}
\end{subfigure}
\hfill
\begin{subfigure}[b]{0.48\textwidth}
         \centering
         \includegraphics[width=\textwidth]{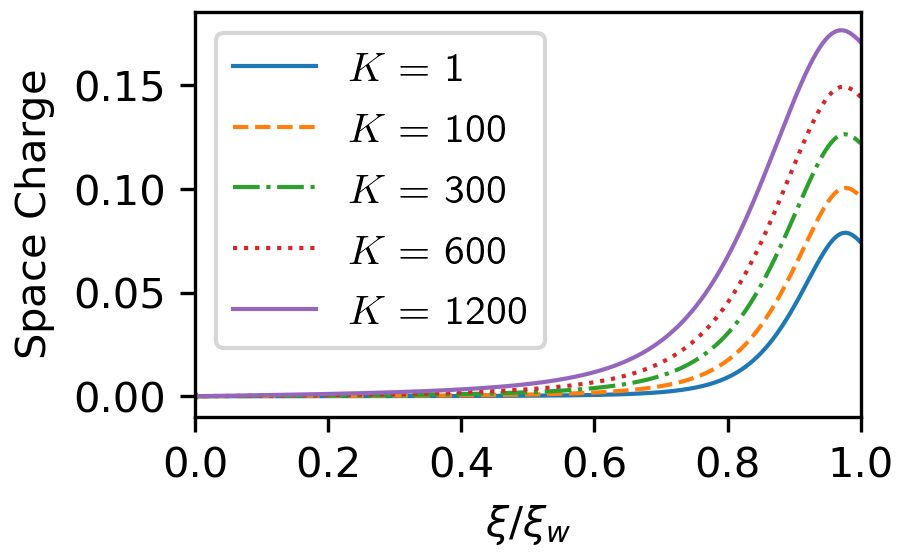}
         \caption{}
\end{subfigure}
\caption{(a) Variation of space charge for collisionless case ($K=1$) with magnetic field $B=0$. (b) Variation of space charge with collision parameter ($K$) for IHM with ${\gamma}_{i0}=1.2$ and ${\gamma}_{e0}=8.8\times10^{4}$}
\end{figure}

\begin{figure}
\centering
\includegraphics{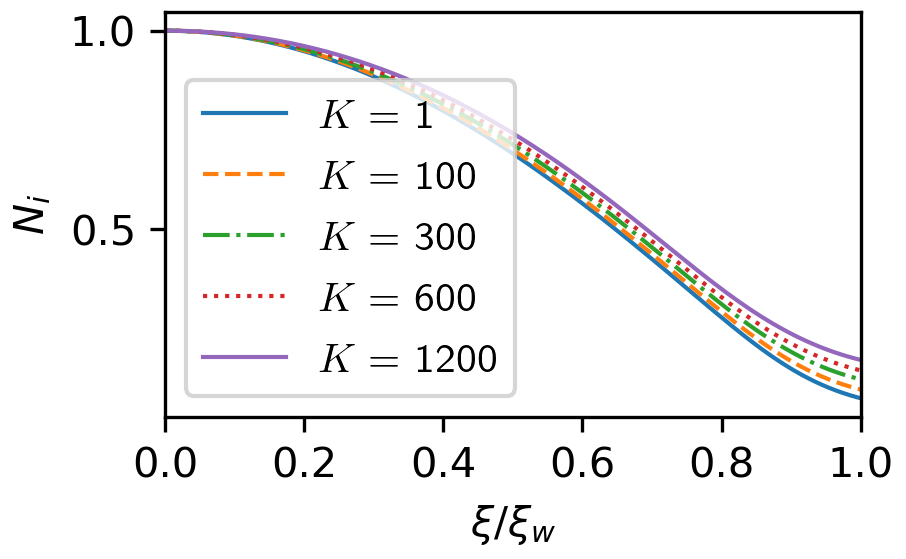}
\caption{Variation of ion density with collision parameter ($K$) for IHM with ${\gamma}_{i0}=1.2$ and ${\gamma}_{e0}=8.8\times10^{4}$}
\end{figure}

\begin{figure}
\centering
\includegraphics{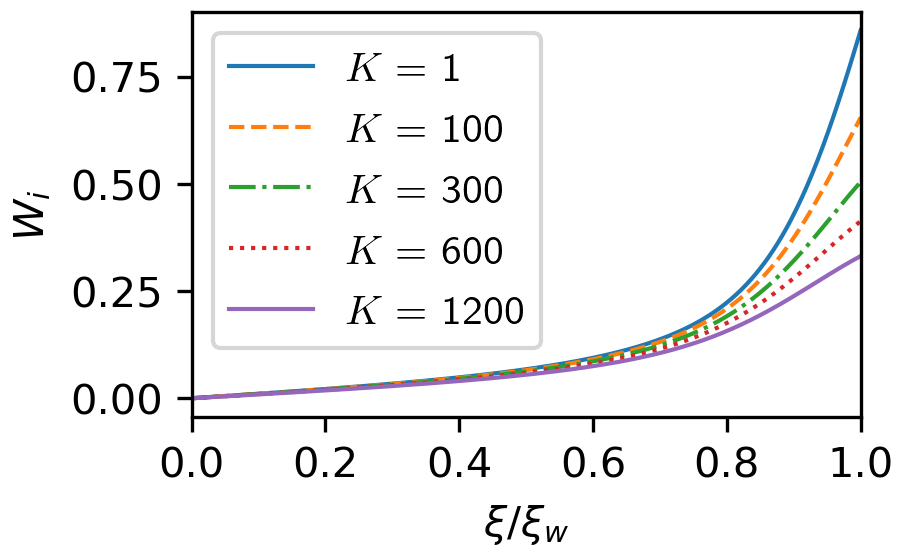}
\caption{Variation of ion velocity along the z-direction with collision parameter ($K$) for IHM with ${\gamma}_{i0}=1.2$ and ${\gamma}_{e0}=8.8\times10^{4}$}
\end{figure}

{Fig. 3} shows the variation of the ion density for different magnitude of collisionality in the system. It is observed that the rate of decrease in ion density remains almost the same for various collision parameter values up to a point ($\xi/{\xi_w}$)$\approx{0.8}$.  After that, for a highly collisional case, the fall of the ion density decreases. This can be explained with the help of {Fig. 4}, \textit{i.e.}, from the velocity component ${w_i}$. The density of ions along the sheath depends upon the z-component of velocity. With the increase of the collision parameter, as we move from the bulk plasma towards the surface, the z-component of velocity up to a point (${\xi}/{\xi_w}$)$\approx{0.8}$ has no significant change. Thereafter, the z-component of velocity falls rapidly with the increase in the collision parameter. Consequently, the ion density inside the sheath falls slowly to conserve the ion flux{\cite{thakur2018}}. It has been observed that the electron density does not respond to ion-neutral collisions in the presence of IHM.

For a particular value of collision parameter, the ion velocity along the z-direction is influenced by three forces,
 \begin{itemize}
     \item Collisional force which always tends to reduce the velocity of ions.
     \item Magnetic force along the negative direction of the z-axis.
     \item Accelerating force towards the surface due to the existing electric field.
 \end{itemize}
However, the magnitude of velocity $w_i$ falls with the increase of the collision parameter. The ion-neutral collision inhibits the ions from moving towards the surface. Therefore, the ions are slowed down, and eventually, they strike the surface at a lower speed.

{Figs. 5} represent the evolution of the electric field. It can be interpreted from the figures that the electric field fall more rapidly with collisions. This can be explained with the help of ion and electron density variation along the sheath length.  The fall of electron density is almost independent of ion-neutral collisions. On the other hand, the collisions, slow down the ions and the ion density falls slowly towards the surface{\cite{thakur2018}}.  This, in turn, increases the space charge and thereby increases the electric field and hence potential.

\begin{figure}
\centering
\includegraphics{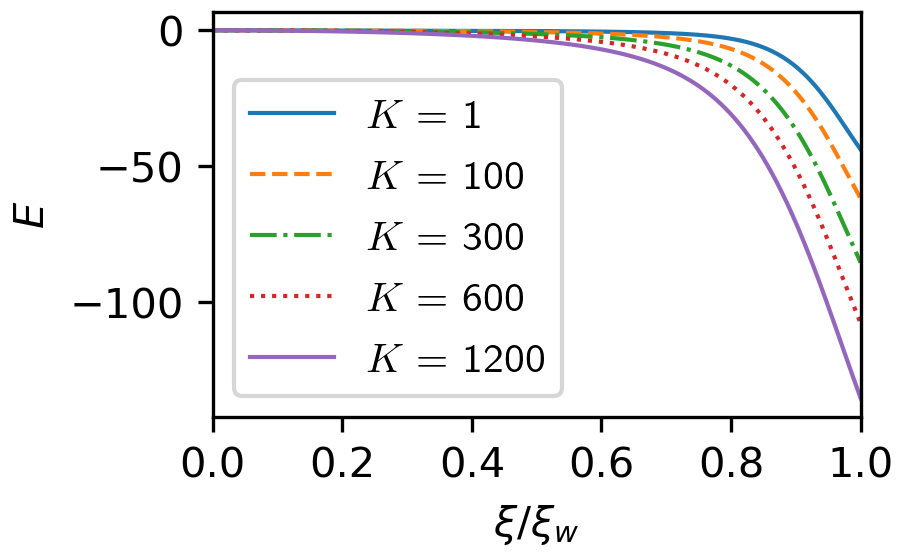}
\caption{Profile of the electric field along the sheath length for various collision parameter ($K$) in the presence of IHM with ${\gamma}_{i0}=1.2$ and ${\gamma}_{e0}=8.8\times10^{4}$}
\end{figure}

\begin{figure}
\centering
\includegraphics{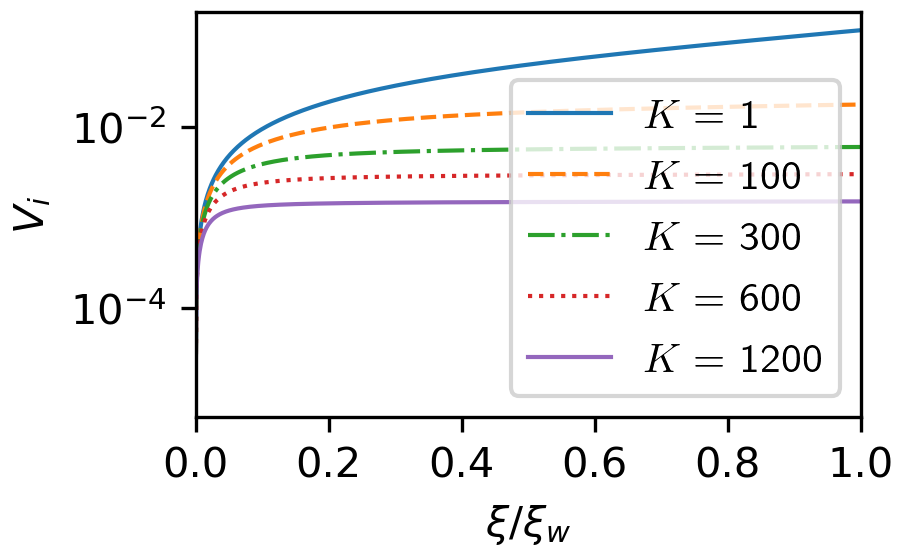}
\caption{Profile of ion velocity  along the y-direction with collision parameter (${\textit{K}}$) for IHM with ${\gamma}_{i0}=1.2$ and ${\gamma}_{e0}=8.8\times10^{4}$}
\end{figure}

\begin{figure}
\centering
\includegraphics{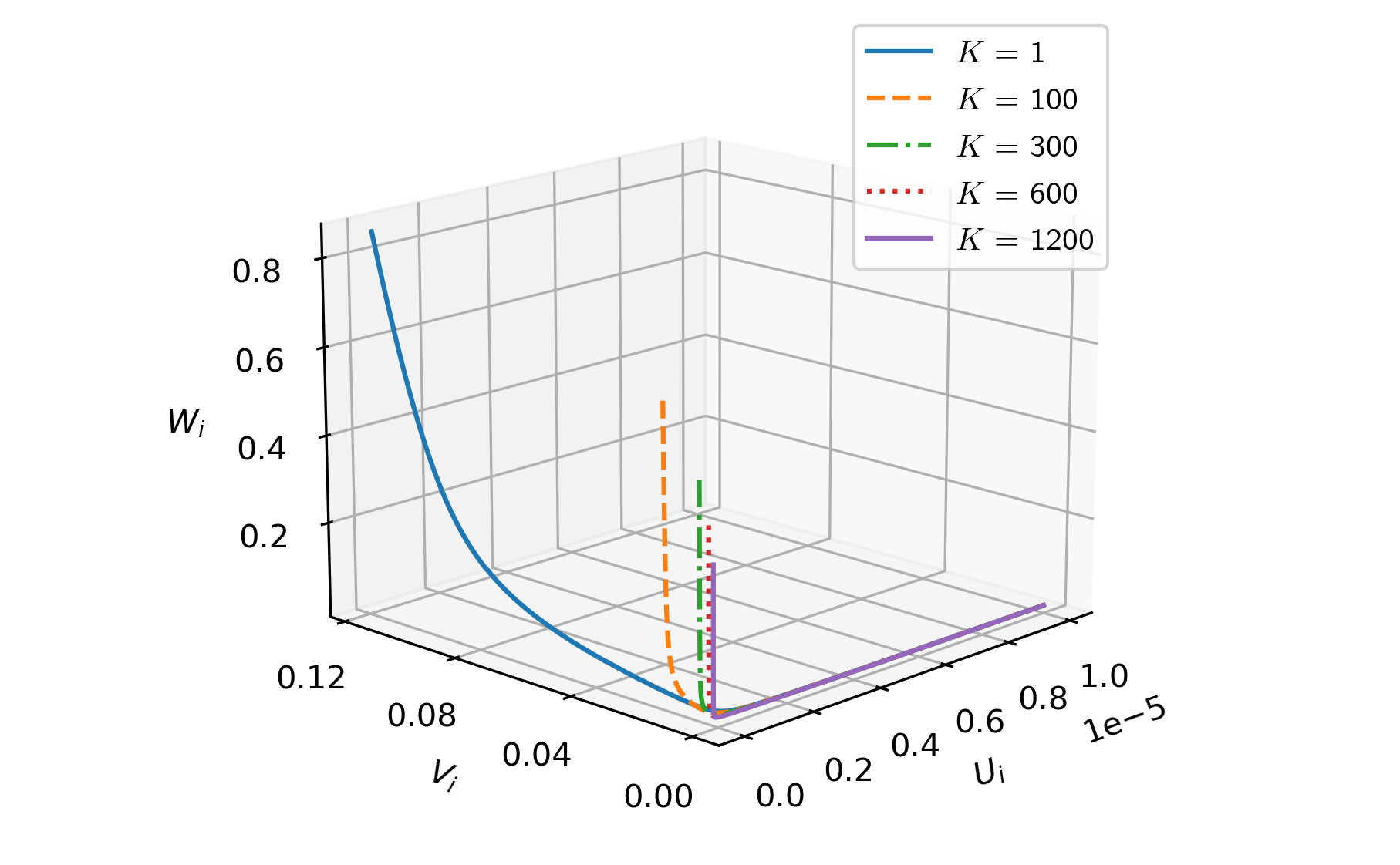}
\caption{3D plot of ion velocity ${u_i}$, ${v_i}$ and ${w_i}$ for various collision parameter ($K$) in the presence of IHM with ${\gamma}_{i0}=1.2$ and ${\gamma}_{e0}=8.8\times10^{4}$}
\end{figure}

\begin{figure}
\includegraphics{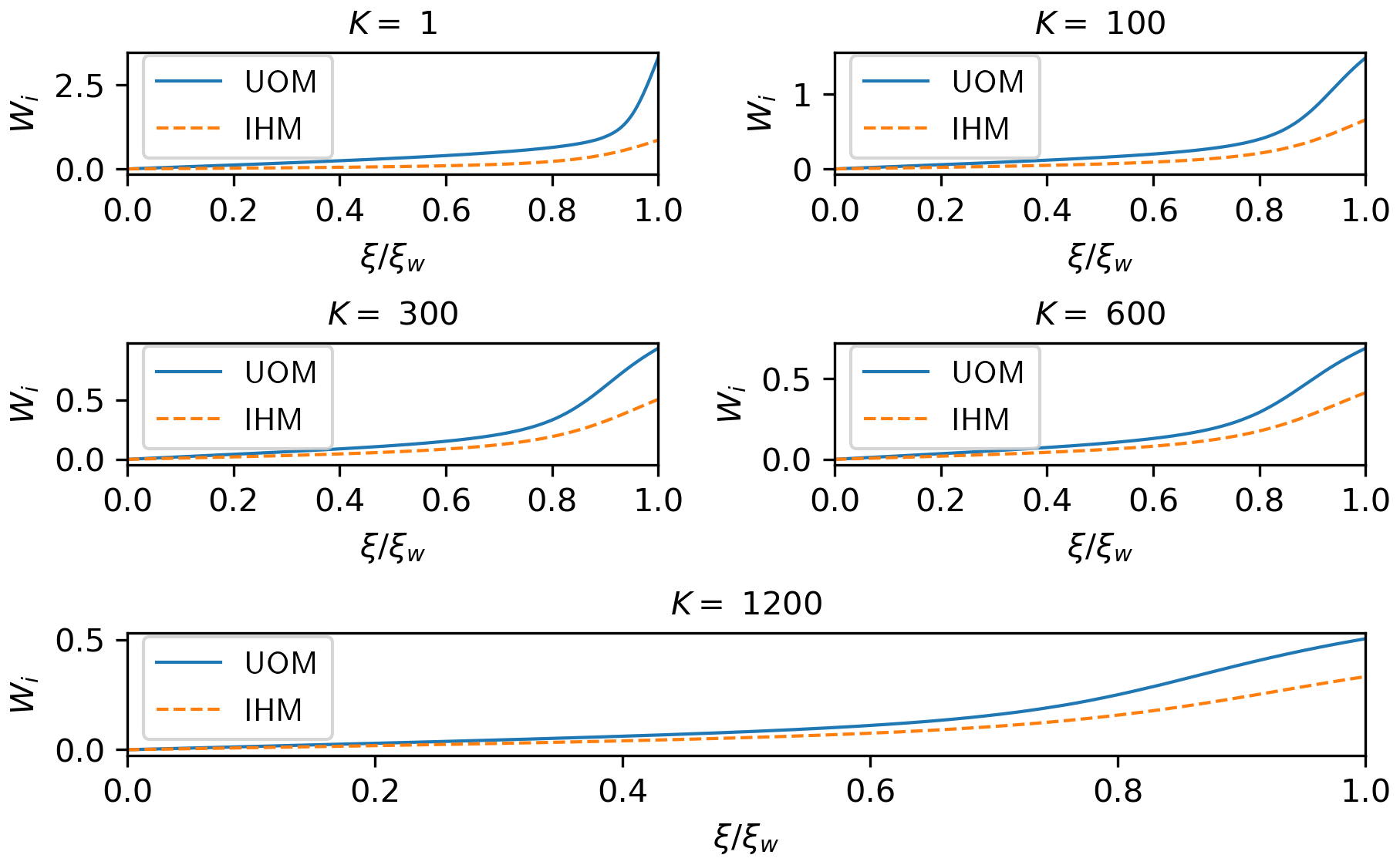}
\centering
\caption{Growth of Ion velocity along the sheath length with collision parameter ($K$) in the presence of UOM (dotted line) and IHM (solid line) with $B_0=0.5T$}
\end{figure}

\begin{figure}
\centering
\includegraphics{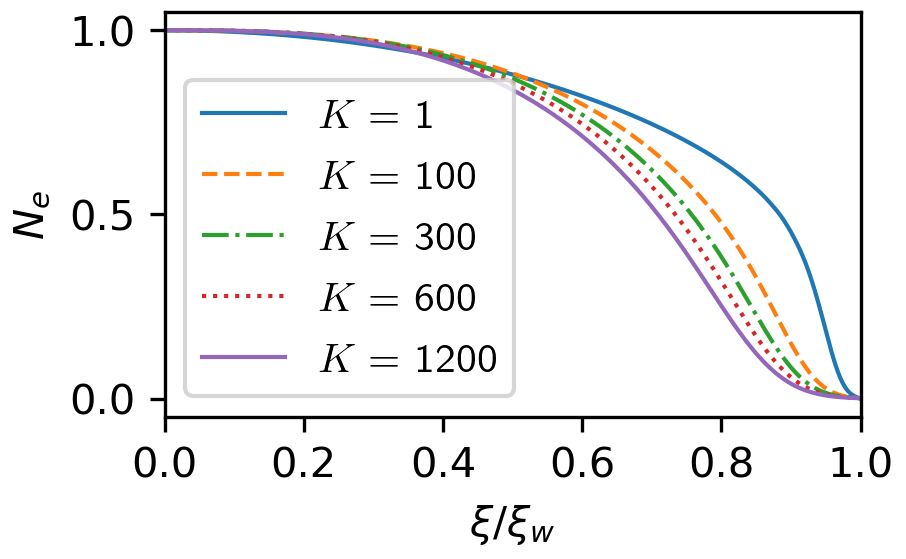}
\caption{Electron density profiles for various values of collision parameter $(K)$ in the presence of UOM field with $B=0.5T$}
\end{figure}

\begin{figure}
\includegraphics{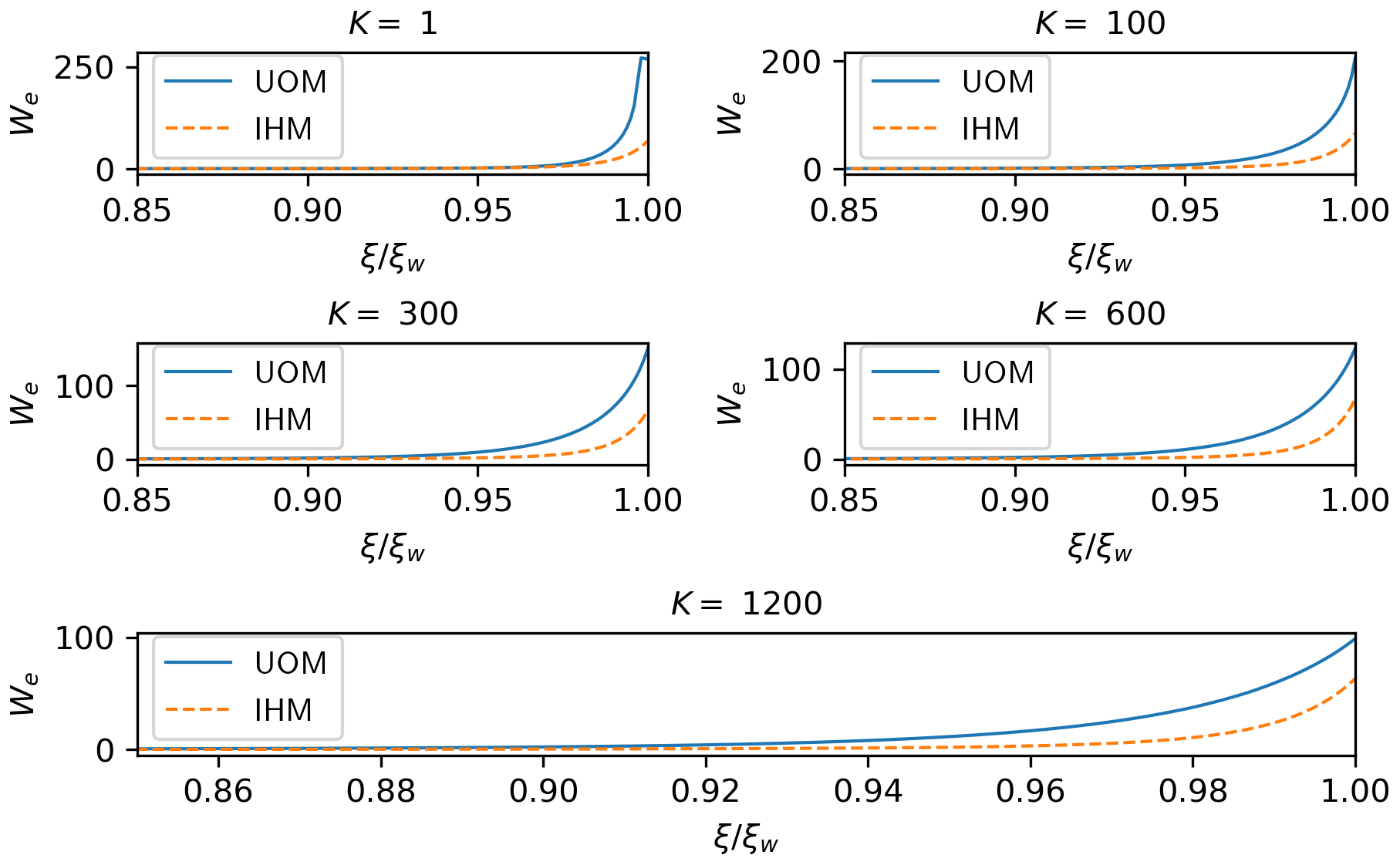}
\centering
\caption{Velocity profiles of electron along the z-direction with collision parameter ($K$) for UOM (dotted line) and IHM (solid line) with $B_0=0.5T$}
\end{figure}

\begin{figure}
\includegraphics{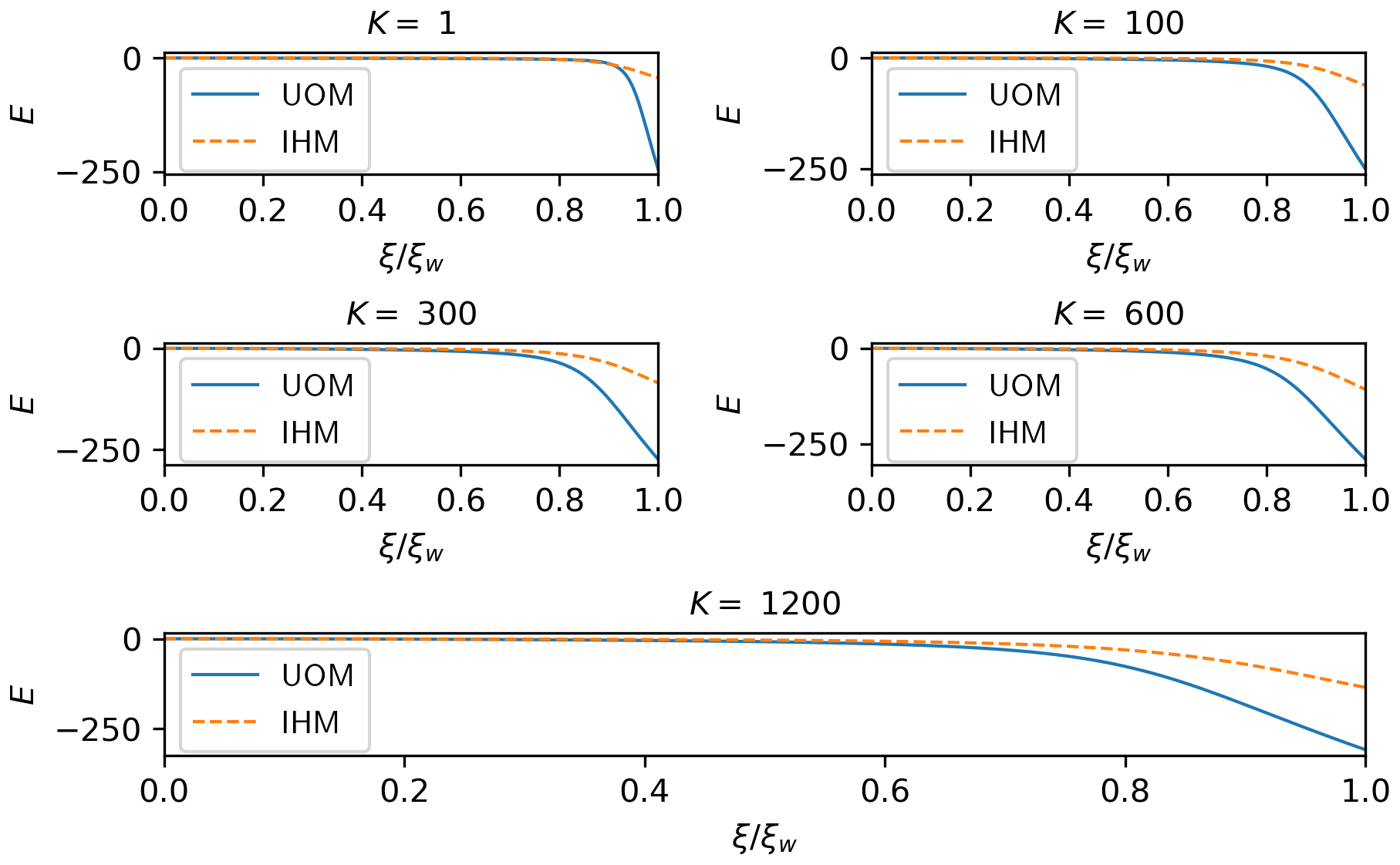}
\centering
\caption{Electric field along the sheath for various values of collision parameter $(K)$ in the presence of UOM (dotted line) and IHM (solid line) with $B_0=0.5T$}
\end{figure}

{Fig. 6} shows the velocity profile along the y-direction. The guiding centre of the gyrating ions has two components of drift velocities along the y-direction. The drift velocities are caused by the presence of the electric field and spatially varying magnetic field{\cite{Chen84}}. They are,
\begin{itemize}
    \item $({v_i})_{E\times B} = \frac{\vec{E}\times\vec{B}}{B^2}$ \textit{i.e.} ${\vec{E}\times\vec{B}}$ drift velocity along the positive y-direction. The ${\vec{E}\times \vec{B}}$ drift velocity of the ions and the electrons are in the same direction{\cite{geraldini2017}}.
    \item $({v_i})_{\nabla B} = (\frac{{v_\bot}r_L}{2})\frac{\vec{B}\times\vec{\nabla}{B}}{B^2}$  \textit{i.e.} ${\vec{\nabla}{B}}$ drift velocity along the negative y-direction.
\end{itemize}

The magnetic force acting along the y-direction increases linearly towards the surface. For a specific value of collision parameter, the magnetic force tends to increase the y-component of velocity, and ion-neutral collisions try to reduce it. The $\vec{E}\times \vec{B}$ drift velocity being in the positive y-direction contributes to the velocity ${v_i}$, and the grad-B drift velocity tends to diminish it. Under the combined effect of all these factors, velocity along the y-direction evolves as depicted in {Fig. 6}. As the ion-neutral collisions slow down the ions, the y-component of the velocity falls with the increase in collisions. At a higher value of collision parameter the velocity component, $v_i$ almost dies out. 
The velocity of ions along the direction of the x-axis is negligible as there is no force except that of the collisional force, which always tends to reduce the velocity of the particles.

 {Fig. 7} shows 3D plot for the evolution of velocity components $u_i,v_i$ and $w_i$ with collision parameter. From the figure, it can be seen that the velocity  $w_i$ dominates over $u_i$ and $v_i$. The dominance becomes more prominent as the velocity profile rises sharply along the z-axis with the increase in collision parameter. Thus, the component $w_i$ gets the highest share of ion velocity. The dominance of $w_i$ over the other two components may be due to the presence of the accelerating electric field along the sheath length. It has been observed that the ion-neutral collisions can not affect the velocity of electrons in the plasma. Therefore, the electron velocities remain unchanged with the increase in ion-neutral collision frequency.

Figs. {8} depict the variation of ion velocity ($w_i$) for UOM and IHM. The ion velocity rises slowly in the presence of the IHM than in comparison to the UOM. But, with the increase in the collision parameter, the difference in the rising rate decreases as a function of distance. The field strength of the IHM increases towards the wall. Therefore, the field provides more constraint to the movement of the ions towards the wall. This causes the velocity to decrease in the presence of the IHM. 

{Fig. 10} shows the variation of electron velocity $w_e$ along the sheath length for two configurations of the magnetic field.  It is seen from the figure that the velocity($w_e$) for the UOM rises rapidly than that of the IHM and there is a significant reduction in the velocity of electrons near the wall in the presence of the IHM.  {Fig. 11} reveals that the presence of IHM parallel to the surface significantly affects the electric field.  {Fig. 9} shows that in the case of the UOM, the electron density declines slightly faster with the increase in the collision parameter.  But, the electron density is almost independent of the collision frequency in the presence of the IHM.  As a result, the space charge will be less in the presence of IHM. This causes the electric field, and consequently the potential to decrease in the presence of the IHM.

\section{Conclusions}
Using a two-fluid hydrodynamic model, the properties of a collisional plasma sheath in the presence of IHM is explored. It has been seen that with the increase in collision parameter or ion-neutral collision frequency,
\begin{itemize}
  \item Plasma sheath expands as the ion-neutral collision restrict the movement of the ions towards the wall. 
  \item   More and more ions begin to deposit with the increase of collision frequency, due to this the space charge increases.
  \item The increase of space charge with collision causes the electric field and thereby the potential to increase.
 \item The $z$ component of ion velocity, $w_i$  dominates over the other components $u_i$ and $v_i$ due to the existing accelerating electric field along z. At higher values of ion-neutral collisions, only the z-component of ion velocity is significant.

\end{itemize}

A comparative study of the plasma sheath parameters has been carried out between our case and UOM(with an angle of inclination, ${\theta=45^{\circ}}$ and the field is lying in the x-z plane). The study reveals that the z-component of ion velocity rises slowly in the presence of the IHM. The IHM parallel to the wall reduces the mobility of ions in the direction perpendicular to the field. Therefore, the ion velocity is always less in the present study than the uniform and oblique magnetic field. Therefore, the kinetic energy of the ions striking the surface decreases. It reduces the heat flux of the ions, and hence erosion of the surface.  

The study has unveiled that the IHM parallel to the surface significantly reduces the value of electric field and potential. This happens because the IHM parallel to the surface restrict the movement of both the electrons and ions towards the wall. Hence, it is expected to enhance the plasma confinement. It is also seen that the electron density is independent of ion-neutral collisions for the IHM, but the density falls rapidly with collisions for the uniform and oblique field.

The magnetic field tends to restrict the ions from moving towards the surface. The IHM with increasing magnitude towards the surface will impose more resistance to the movement of the ions in the direction of the surface. The ion-neutral collision plays a role similar to that of the magnetic field. Therefore, the combination of ion-neutral collision and IHM may be useful in confining a plasma. Further, the flow of ions towards the surface may be controlled by optimizing the frequency of collision, and the configuration of the IHM. The future aspect of the present study is to develop an experimental model for it, which is supposed to help in the plasma processing of materials. The study of ion dynamics may help to develop the model. However, a more detailed analysis is required for a better understanding of the problem.
\section*{Appendix: Mathematical Description of Magnetic field profile}
In this section, the magnetic field profile has been discussed. The magnetic field profile is represented as

\begin{equation}\vec{B}=B_0(1+{\alpha}z)\hat{i}+(B_0{\alpha}x)\hat{k}\tag{A1}\end{equation}

The first term of equation (A1) is called the gradient term which is the prime concern of this study. The second term is related to the curvature of the field lines and is called the curvature term.

Now, the components of magnetic forces acting on electrons and ions are as follows, 
\begin{equation}
e(\vec{v_i}{\times}\vec{B})=eB_0{\alpha}x{v_{iy}}\hat{i}+eB_0[v_{iz}(1+{\alpha}z)+B_0{\alpha}xv_{ix}]\hat{j}-ev_{iy}B_0(1+{\alpha}z)\hat{k}\tag{A2}
\end{equation}

\begin{equation}
\begin{multlined}
e(\vec{v_e}{\times}\vec{B})=eB_0{\alpha}x{v_{ey}}\hat{i}+eB_0[v_{ez}(1+{\alpha}z)+B_0{\alpha}xv_{ex}]\hat{j}-ev_{ey}B_0(1+{\alpha}z)\hat{k}\tag{A3}
\end{multlined}
\end{equation}
Where, $\hat{i}$, $\hat{j}$ and $\hat{k}$ represent the unit vectors along x, y, and z axes, respectively.

In order to incorporate the effect of only the gradient term  (which dominates over the curvature term) of the magnetic field, the study of sheath is performed along the z-axis \textit{i.e.} $x=0$ line. Now, putting $x=0$ in equations (A4) and (A5), the components of forces takes the form,

\begin{equation}
e(\vec{v_i}{\times}\vec{B})= eB_0v_{iz}(1+{\alpha}z)\hat{j}-ev_{iy}B_0(1+{\alpha}z)\hat{k}\tag{A4}
\end{equation}

\begin{equation}
e(\vec{v_e}{\times}\vec{B})= eB_0v_{ez}(1+{\alpha}z)\hat{j}-ev_{ey}B_0(1+{\alpha}z)\hat{k}\tag{A5}
\end{equation}

Thus, it is seen that the curvature term does not affect the sheath formation along z-direction despite its presence in the original magnetic field. Hence, the magnetic field profile for the present study can be taken as,

\begin{equation}
\vec{B}=B_0(1+{\alpha}z)\hat{i} \tag{A6}
\end{equation}

\section*{DATA AVAILABILITY STATEMENT}
The data that support the findings of this study are available from the corresponding author upon reasonable request.

\end{document}